\documentclass[sigconf]{cidr-2025}
\usepackage{subfig}
\AtBeginDocument{%
  }
\setcopyright{acmlicensed}
\copyrightyear{2025}
\acmYear{2025}
\acmBooktitle{}
\usepackage{xspace}
\newcommand{\PaperAcronym}{DataLens\xspace}
\newcommand{\quotes}[1]{``#1''}
\begin{document}
\title{\PaperAcronym: \\ML-Oriented Interactive Tabular Data Quality Dashboard}
\author{Mohamed Abdelaal}
\affiliation{%
	\institution{Software AG, Darmstadt, Germany}
	\streetaddress{Uhlandstrasse 12}
	\city{}
	\country{}
}
\email{Mohamed.Abdelaal@softwareag.com}

\author{Samuel Lokadjaja}
\affiliation{%
	\institution{TU Darmstadt, Darmstadt, Germany}
	\city{}
	\country{}
}
\email{sammyloka@yahoo.com}
\author{Arne Kreuz}
\affiliation{%
	\institution{Software AG, Darmstadt, Germany}
	\city{}
	\country{}
}
\email{Arne.Kreuz@softwareag.com}

\author{Harald Schöning}
\affiliation{%
	\institution{Software AG, Darmstadt, Germany}	\city{}
	\country{}
}
\email{Harald.Schoening@softwareag.com}
\renewcommand{\shortauthors}{Abdelaal et al.}

\begin{abstract}
Maintaining high data quality is crucial for reliable data analysis and machine learning (ML). However, existing data quality management tools often lack automation, interactivity, and integration with ML workflows. This demonstration paper introduces \PaperAcronym{}\footnote{A recording of the demonstration can be found at \url{https://youtu.be/tW5qqFqDFYI}.}, a novel interactive dashboard designed to streamline and automate the data quality management process for tabular data. \PaperAcronym{} integrates a suite of data profiling, error detection, and repair tools, including statistical, rule-based, and ML-based methods. It features a user-in-the-loop module for interactive rule validation, data labeling, and custom rule definition, enabling domain experts to guide the cleaning process. Furthermore, \PaperAcronym{} implements an iterative cleaning module that automatically selects optimal cleaning tools based on downstream ML model performance. To ensure reproducibility, \PaperAcronym{} generates DataSheets capturing essential metadata and integrates with MLflow and Delta Lake for experiment tracking and data version control. This demonstration showcases \PaperAcronym{}'s capabilities in effectively identifying and correcting data errors, improving data quality for downstream tasks, and promoting reproducibility in data cleaning pipelines. 
\end{abstract}

\maketitle
%
%
\section{Introduction}\label{sec:intro}
%
\begin{figure*}
    \centering
    \includegraphics[trim=0.5cm 0.8cm 0.5cm 0.5cm,clip,width=0.85\linewidth]{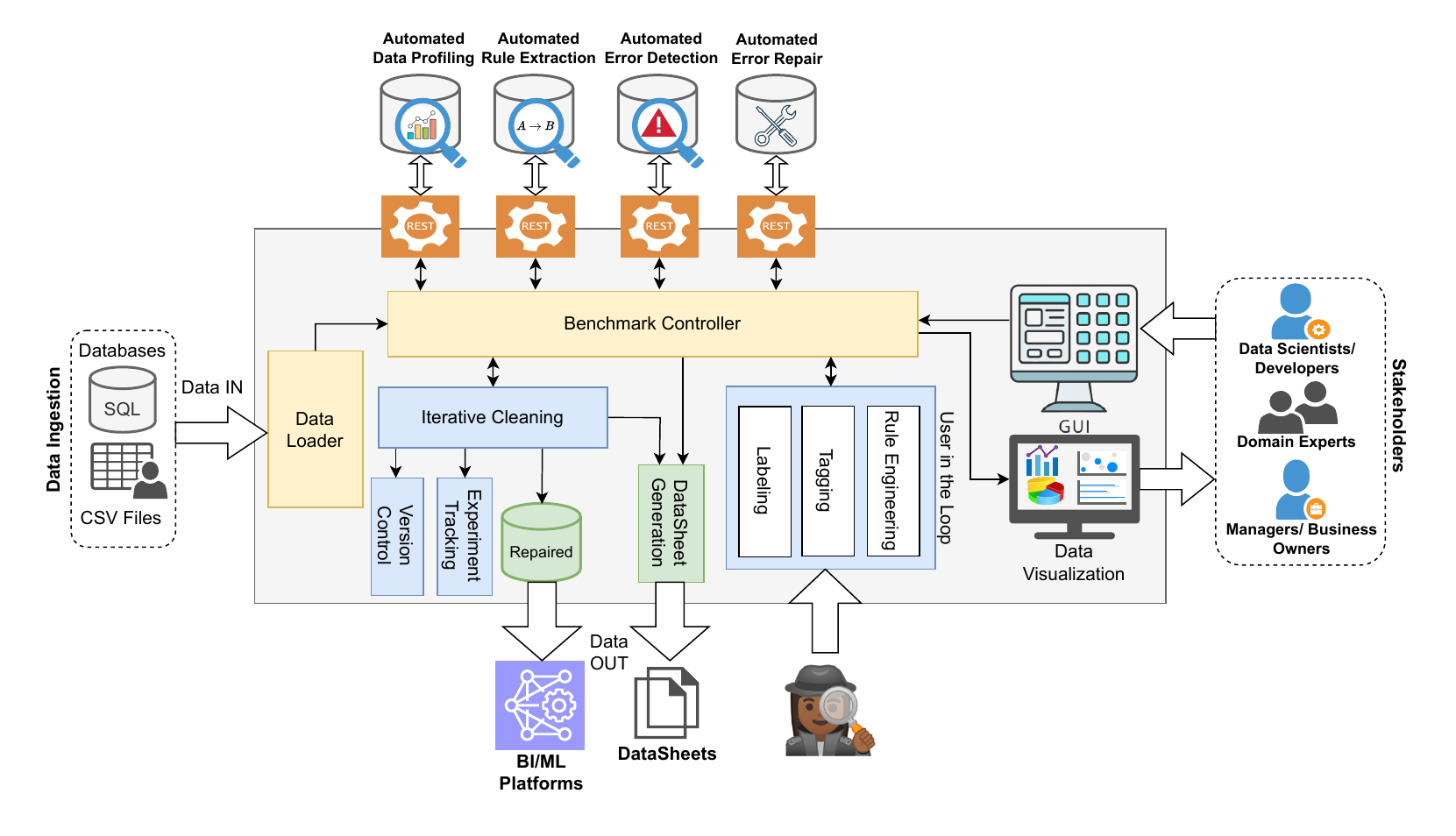}
    \caption{Architecture of \PaperAcronym}\vspace{-3mm}
    \label{fig:architecture}
\end{figure*}

Data quality is critical in data analysis pipelines, directly impacting the reliability and validity of insights derived from that data. A recent study by Gartner found that poor data quality costs organizations an average of \$12.9 million per year~\cite{McKinsey2016Analytics}. Inaccurate or incomplete data can lead to faulty decisions and inaccurate predictions, negatively affecting strategic planning and operational efficiency. For instance, missing or incorrect values can cause machine learning (ML) models to produce unreliable results. Inconsistent data, in which the same item is represented differently in distinct records, can disrupt the analysis process and lead to incorrect conclusions. Furthermore, outdated or irrelevant data can skew the overall understanding of business performance, leading to misguided strategies. Accordingly, high data quality is essential for businesses to make informed decisions and maintain competitiveness.

Data quality management typically involves several processes, including data profiling and data cleaning. The former is the process of examining the data and collecting statistics, rules, or informative summaries about that data. Data cleaning involves identifying and correcting corrupt or inaccurate records. Techniques used in this process can range from simple steps like removing duplicates or filling missing values, to more complex operations like using ML-based methods to identify outliers and decide how to handle them. Crafting a data quality management pipeline can be challenging without broad expertise in data science for several reasons. First, there is a vast array of tools and technologies available, each with its unique strengths and limitations. Knowing which tool to use, when to use it, and how to best use it requires a deep understanding of both the tools themselves and the specific data quality issues at hand. Second, effective data quality management requires not just the ability to apply tools, but also the knowledge to understand the implications of the quality issues on the data pipeline and downstream tasks. Additionally, maintaining data quality is not a one-time task. It requires ongoing monitoring and adjustment as new data comes in, as the nature of the data changes, or as the business requirements evolve.  

To overcome these challenges, we introduce in this paper an interactive data quality dashboard, referred to as \PaperAcronym{}\footnote{Source code will be publicly available with the camera-ready version of the paper.}, designed to streamline and automate multiple aspects of the data quality management process. With its modular design, \PaperAcronym{} brings significant enhancements to the field of data science, particularly in data profiling, validation, error detection, and correction. To this end, \PaperAcronym{} integrates several data preparation tools via REST APIs, facilitating the extension of the dashboard's capabilities. For instance, \PaperAcronym{} includes data profiling tools that automatically generate insights and rules, aiding users in exploring the data effectively. Moreover, \PaperAcronym{} integrates statistical, rule-based, and ML-based error detection tools. For the ML-based tools, the dashboard enables users to label data instances, a required step while training detection models. \PaperAcronym{} also integrates advanced algorithms to propose and apply corrections to identified data errors, reducing the manual intervention traditionally required in data cleaning processes. 

\PaperAcronym{} integrates with ML tracking tools like MLflow for seamless tracking of data quality experiments, models, and results, offering a unified view of data quality and ML processes. It also supports Delta Lake integration, enabling version control of data, enhancing robustness, and allowing rollback to previous data versions. Additionally, \PaperAcronym{} generates DataSheets, JSON files that capture critical information including data version tags, hyperparameters, generated rules, data quality metrics, employed cleaning tools, and more. The dashboard offers two important features, including the user-in-the-loop module and the iterative cleaning module. The former module facilitates user interaction with the dashboard to tag dirty values, label data instances, and add custom rules. The latter module automatically selects the optimal data-cleaning tools that improve downstream ML models.  Thanks to these features, the role of the data users is streamlined and focused, limited to validating the generated rules and labeling data samples to train ML models used for error detection and/or correction. This approach maximizes the value of domain expert input while minimizing the technical burden on them.

To sum up, this paper presents several key contributions: (1) a modular design for a data quality dashboard that enhances its extensibility; (2) the formulation of data cleaning tool selection as a hyperparameter tuning problem, allowing for automatic pipeline configuration based on downstream ML model performance; (3) the generation of DataSheets to precisely track the data cleaning process and ensure reproducibility; (4) the implementation of a user-in-the-loop module for interactive data preparation; (5) providing a means to interactively collect user labels for training ML-based error detection tools. This is a significant departure from the norm, where developers of such tools typically rely on ground truth data for labels. By enabling user-driven labeling, \PaperAcronym{} offers a more realistic evaluation framework for ML-based error detection tools; and (5) enabling the execution of multiple error detection tools, with \PaperAcronym{} autonomously integrating and deduplicating results to improve the precision and recall of error detection. To the best of our knowledge, \PaperAcronym{} is the first dashboard to offer interactive, automated, and ML-oriented data quality management.  



%
\vspace{-2mm}\section{Architecture \& Overview}\label{sec:overview}
%
\sloppy In this section, we introduce the architecture of \PaperAcronym, depicted in Figure~\ref{fig:architecture}. The architecture constitutes an automated, interactive data quality dashboard designed to optimize data quality for downstream applications such as Business Intelligence (BI) and ML platforms. The process begins with data ingestion, where data can be ingested into \PaperAcronym{} via one of three methods: (1) using one of the preloaded datasets that come with the dashboard, allowing users to explore its functionalities without needing their data; (2) uploading CSV or Excel files; or (3) establishing a SQL database connection. When a dataset is uploaded as a CSV or Excel file, an automatic backend process is initiated. A dedicated folder, named after the uploaded file, is created to store the dataset as 'dirty.csv'. A subfolder within this directory is also created to house the Delta table associated with this dataset. Additionally, the uploaded dataset is stored in the backend as a pandas DataFrame. For database connections, users can connect the dashboard to MySQL, PostgreSQL, and Microsoft SQL Server databases. They can input their credentials and specify the table and dataset they wish to load. Once loaded, these tables are treated identically to uploaded files.

\PaperAcronym{} incorporates a data loader that feeds the input data into a dashboard controller. Such a controller is mainly responsible for controlling other modules, regulating data flow, and ensuring each step is executed properly. For a modular design, \PaperAcronym can integrate with several external tools using a set of standard REST APIs. An automated data profiling module analyzes the ingested data, identifying and recording its characteristics. Concurrently, an automated rule extraction module generates rules used later while detecting and repairing data errors. This automated rule extraction leverages advanced algorithms, considering both the statistical properties and domain-specific characteristics of the data. The automated error detection module implements statistical, rule-based, and ML-based tools that scan the data to identify inconsistencies, outliers, or other potential issues. Identified errors are then passed to the automated error repair module, which leverages advanced algorithms to propose and apply corrections to the detected errors. In parallel, a version control tracking module maintains a record of each data version throughout the cleaning process. This allows for robust data management and facilitates rollback to previous data versions if necessary. 

\PaperAcronym{} is enhanced by two crucial features: the \textit{user-in-the-loop} module and the \textit{iterative cleaning} module. The user-in-the-loop module supports active user involvement, empowering them to validate or adjust the system-generated rules and corrections, as well as to introduce their own custom rules. It also enables users to annotate specific data samples to train ML models utilized for data validation or correction. Furthermore, this module facilitates proactive error management by allowing users to tag data samples known to be corrupted in advance. The iterative cleaning module autonomously identifies the optimal error detection and repair tools for specific input data used in training a downstream ML model. To this end, we conceptualize the selection of error detection and repair tools as a hyperparameter tuning problem. By conducting multiple cleaning iterations with various combinations of these tools, the iterative cleaning module determines the configuration that maximizes predictive accuracy.

The system also generates \textit{DataSheets} capturing critical information such as data version tags, hyperparameters, generated rules, data quality metrics, and employed cleaning tools. Another output of \PaperAcronym is the repaired data, which can be utilized in BI or ML platforms. Additionally, the system offers visualizations that allow various stakeholders—including data scientists, developers, domain experts, managers, and business owners—to review and comprehend the data cleaning process and its outcomes. \PaperAcronym{} has been designed with integration capabilities with two common ML tracking tools, such as MLflow. This allows for seamless tracking of data quality experiments, models, and results, providing a unified view of both data quality management and ML processes. \PaperAcronym{} also includes integration with Delta Lake, enabling tracking of different data versions. This feature brings robustness to the data management process, allowing the tracking of changes over time, and facilitating rollback to previous data versions if required.

\begin{figure*}
    \centering
    \includegraphics[width=0.97\linewidth]{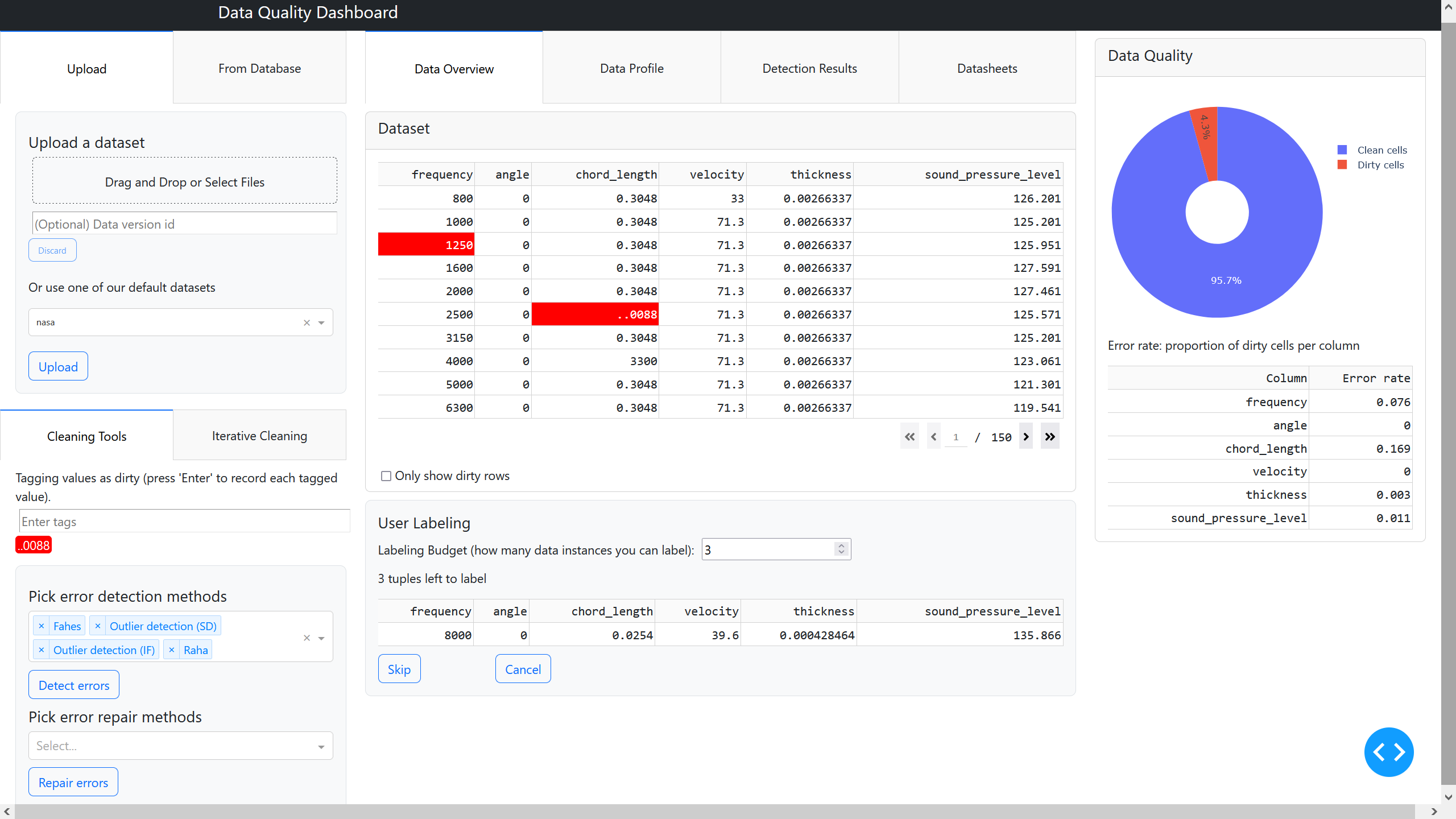}
    \caption{Main window of \PaperAcronym}
    \label{fig:main_window}\vspace{-2mm}
\end{figure*}

Figure~\ref{fig:main_window} depicts the main window of \PaperAcronym. The left segment of the dashboard contains a user interface for data upload. Users can also select from a range of automated error detection and repair tools, enhancing the adaptability of the system to diverse data quality needs. The central part of the dashboard is organized into several tabs: Data Overview, Data Profile, Error Detection Results, and DataSheets. The Data Overview tab displays both the uploaded datasets and any detected errors, providing a quick snapshot of the data and its quality issues. A dedicated user labeling section is included here, where users can label selected data samples as either 'true' (i.e., dirty) or 'false' (i.e., clean). The Data Profile tab presents the findings of the data profiling tools, offering insights into the characteristics and structure of the dataset. The Error Detection Results tab, meanwhile, visualizes the output of the error detection tools, making it easy to understand and manage detected errors. The DataSheets tab displays the generated DataSheet with important metadata and gives users the functionality to download these files for further analysis or record-keeping. Finally, the right segment of the dashboard hosts the Data Quality section, providing visual representations of various data quality metrics.
%
\vspace{-4mm}\section{Data Quality Tools}\label{sec:tools}

To enhance data quality, we opted to integrate a suite of existing, readily available tools. REST API serves as a standardized interface that facilitates interaction between \PaperAcronym{} and such external data quality tools. These APIs enable the exchange of data and instructions, ensuring seamless integration and operation within the system. This uniform communication enhances the system's scalability and flexibility, allowing for the straightforward incorporation of new tools and services as needed. We have implemented the REST APIs using the FastAPI framework. \PaperAcronym{} includes several API calls to facilitate communication with external tools, such as POST, GET, and PUT. The POST method forwards tasks or requests to external tools, the GET method retrieves results from these tools, and the PUT method updates information related to specific requests.

\textbf{Automated Data Profiling:} The \quotes{Data Profile} tab provides a detailed overview of the dataset's attributes, ensuring users have immediate access to vital information. This includes functional dependency (FD) rules and a comprehensive data profile. \PaperAcronym{} leverages the YData-profiling library~\cite{YDataProfiling} to generate an extensive data profile report. This report provides a thorough overview of the dataset, including descriptive statistics, visualizations of data distribution for each column, measures of central tendency, and variable frequencies. Additionally, it identifies correlations between variables, detects missing data points, and flags potential data quality issues. This holistic data profiling approach aids users in gaining a deeper understanding of the dataset's characteristics.

To automatically identify FD rules, \PaperAcronym{} utilizes the command-line interface of Metanome~\cite{Papenbrock2015}, along with FD detectors such as HyFD~\cite{Papenbrock2016} and Tane~\cite{Huhtala1999}. \PaperAcronym{} empowers users to validate automatically generated FD rules and \textbf{engineer custom rules}, ensuring accuracy and comprehensiveness. While automated rule generation offers a valuable starting point, it may produce irrelevant or imprecise rules. For instance, tools might incorrectly infer a relationship between a district's zip code and the number of inhabitants. \PaperAcronym{} allows users to review, confirm, modify, or reject these automatically generated rules. Furthermore, users can define additional rules not captured by automated methods. To define a new rule, users must specify at least one column as the determinant (i.e., the column on which the dependency relies) and at least one column as the dependent (i.e., the column affected by the determinant). This straightforward mechanism allows users to easily incorporate their domain knowledge and specific data requirements into the system.

\textbf{Automated Error Detection}: \PaperAcronym{} incorporates a diverse set of error detection tools, ranging from statistical methods like standard deviation (SD), interquartile range (IQR), and isolation forest (IF) for outlier detection, to specialized tools like FAHES for disguised missing data detection, NADEEF for rule-based error detection, and KATARA for knowledge-based error detection. It also includes probabilistic data repair tools like HoloClean, missing value detection tools like MV Detector, ML-based error detection tools like RAHA, and ensemble methods like Min-K, which combines the detections of multiple methods. A detailed explanation of these tools can be found in \cite{rein23}. Recognizing that different tools excel at detecting different error types, \PaperAcronym{} allows users to select multiple tools for execution. These tools are executed sequentially in the backend, and \PaperAcronym{} automatically consolidates their detections into a single array, filtering out duplicates. 

\begin{figure}[htbp] 
	\centering
    \subfloat[NASA]{\label{fig:nasa_raha}\includegraphics[width=0.5\linewidth]{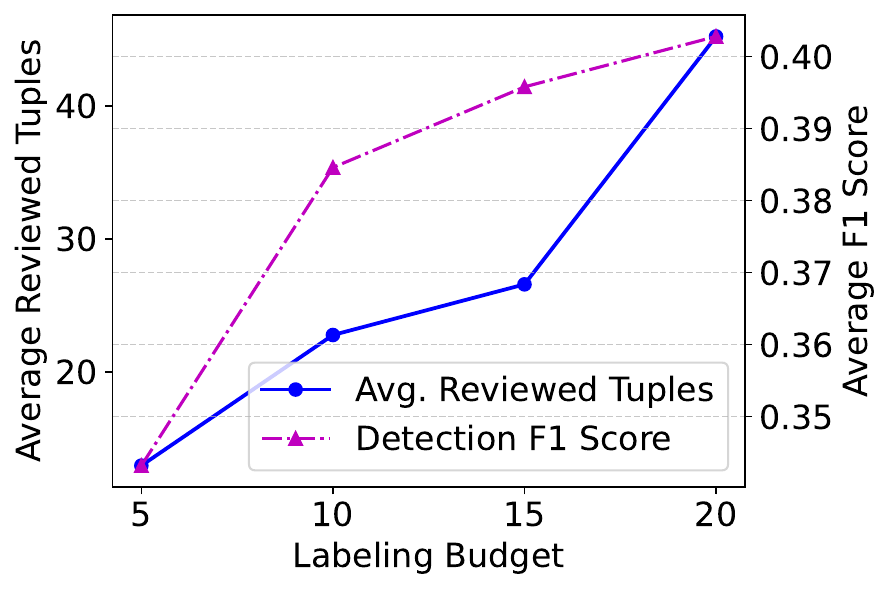}}	%
	\subfloat[Beers]{\label{fig:beers_raha}\includegraphics[width=0.5\linewidth]{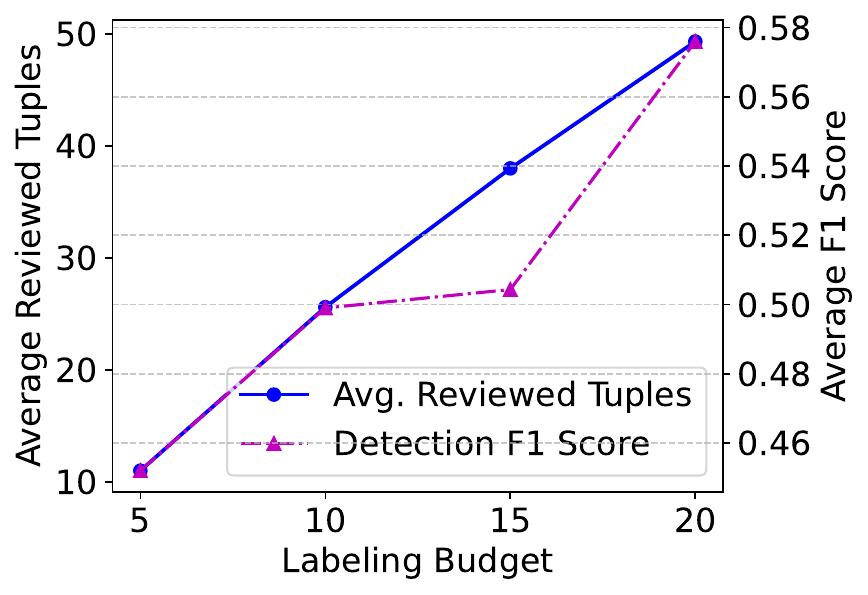}}
	\caption{Evaluation of labeling ML-based tools}
 	\label{fig:labeling_results} \vspace{-3mm}
 \end{figure}

RAHA employs a user-dependent operational modality, unlike other detection methods that operate independently of user input. While RAHA is initiated simultaneously with other detection methods, its results computation and visualization occur asynchronously, contingent upon the user completing the required tuple labeling. This design ensures that RAHA's output accurately reflects the user's input, preventing the premature display of results before data labeling is complete. To facilitate \textbf{user labeling}, \PaperAcronym{} prompts the user to define a labeling budget ($N$), representing the number of tuples they are willing to label. Subsequently, the dashboard presents $N$ tuples sequentially for labeling. The user examines each tuple, marking any dirty instances. If a tuple contains no dirty instances, the user can skip it, prompting \PaperAcronym{} to display the next recommended tuple. As a tuple selection strategy, RAHA leverages clustering with label propagation. Figure \ref{fig:labeling_results} presents an evaluation of the data labeling process for RAHA, highlighting a key advantage of \PaperAcronym{} in facilitating realistic assessments of ML-based cleaning tools. Unlike evaluations presented in the original publications of such tools, \PaperAcronym{} allows us to quantify the actual labeling effort required. As depicted in Figure \ref{fig:nasa_raha}, the average number of tuples reviewed by users consistently exceeds twice the user-defined budget. For instance, with a budget of 20 tuples, users reviewed an average of 45.2 tuples. This discrepancy arises because the tuple selection strategy, while designed to prioritize potentially erroneous data, often selects clean tuples for review. The figure also demonstrates that increasing the budget from 5 to 20 tuples yields a marginal improvement in the average F1 score for error detection, rising from 0.34 to 0.4. Similar trends are observed for the Beers dataset, as shown in Figure \ref{fig:beers_raha}. 


%

\begin{figure}
    \centering
    \includegraphics[trim=0.5cm 0.6cm 0.5cm 0.5cm,clip,width=1\linewidth]{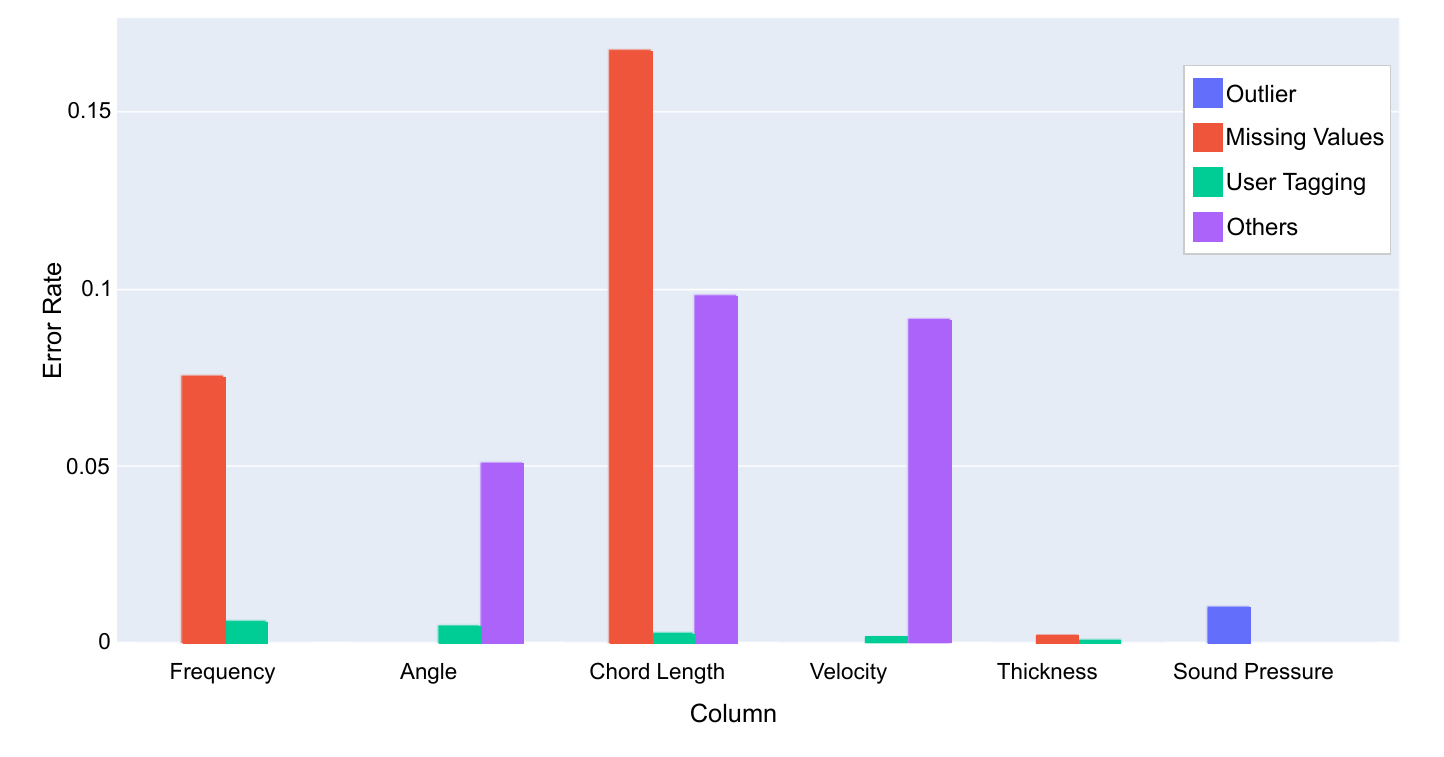}
    \caption{Distribution of detections across various attributes of the NASA dataset}
    \label{fig:detections} \vspace{-3mm}
\end{figure}

Aside from the automated tools, \PaperAcronym{} empowers users to leverage their domain expertise through \textbf{data tagging}. Users can flag known erroneous values (e.g., -1, 0, 99999) within the dataset. The dashboard provides an intuitive interface for submitting these values, which \PaperAcronym{} then incorporates as supplementary labels for ML-based error detection tools. This user-provided knowledge enhances the models' robustness and accuracy. Additionally, \PaperAcronym{} performs a comprehensive search for these tagged values within the dataset, appending their indices to the detection list. This combined approach streamlines the detection of dirty data by leveraging automated tools and user expertise. These user-driven features--data tagging, FD rule engineering, and tuple labeling--collectively constitute the user-in-the-loop module of \PaperAcronym{}. Figure~\ref{fig:detections} visualizes the distribution of detected errors across attributes within the NASA dataset, categorized by error types identified by automated detection tools (IQR, SD, FAHES, and RAHA) or tagged by the user. This visualization, automatically generated by \PaperAcronym{} upon completion of the error detection phase, is presented to the user in the "Detection Results" tab, providing a comprehensive overview of the dataset's error profile.
 
\textbf{Automated Data Repair}: After the error detection phase, users can proceed to error repair. \PaperAcronym{} offers two distinct repair strategies: ML-based and standard imputation. For ML-based imputation, the system employs Decision Tree algorithms for numerical columns and k-nearest Neighbors (k-NN) for categorical columns. In contrast, standard imputation utilizes simpler techniques: the arithmetic mean for numerical columns and a predefined \quotes{Dummy} value for categorical columns. Following repair, the processed data is stored as a CSV file in the input dataset's directory and committed as a new version to the dataset's Delta Lake. This dual storage mechanism ensures version control and traceability and facilitates analysis across different dataset versions over time.

%
\section{Iterative Cleaning}\label{sec:iter_clean}
%
In this section, we introduce the Iterative Cleaning module, designed to automatically select data cleaning tools in light of the performance of a downstream ML model. By tailoring the cleaning process to the specific needs of the ML model, we can effectively optimize the model's performance. Once a user provides the type of ML model to be trained, the Iterative Cleaning module embarks on a cycle of cleaning processes, employing a variety of cleaning tools. The cleaning tools that contribute to the most significant performance enhancement are selected and applied. 
In this way, data-cleaning tools are conceptualized as hyperparameters that can be optimized jointly with the typical parameters in ML pipelines, e.g., number of layers, and number of neurons. 
 
We commence the optimization process by delineating the search space, encompassing all potential permutations of error detection and repair tools. Subsequently, a scoring function is defined to measure the performance of a given detector and repair tool on the repaired dataset. Such a function can be defined as the Mean Square Error (MSE) of the target ML model in case of regression and the F1 score of the ML model in case of classification. An iterative process continues for a predetermined number of iterations, or until the accuracy of the ML model reaches a desired threshold. In each iteration, the module first trains the ML model on a repaired version of the dataset, and calculates the model accuracy. \PaperAcronym{} leverages a Bayesian hyperparameter optimization algorithm, referred to as Optuna~\cite{Akiba2019}, that facilitates the identification of optimal hyperparameters for ML models. Optuna systematically navigates this search space to ascertain the amalgamation of tools that yield the highest performance according to the predefined scoring function. A sequential model-based optimization approach drives this process. Optuna iteratively selects the most promising hyperparameters to evaluate, based on the past trial outcomes, to converge on the optimal configuration efficiently.

\begin{figure}[htbp] 
	\centering
    \subfloat[NASA]{\label{fig:nasa}\includegraphics[width=0.5\linewidth]{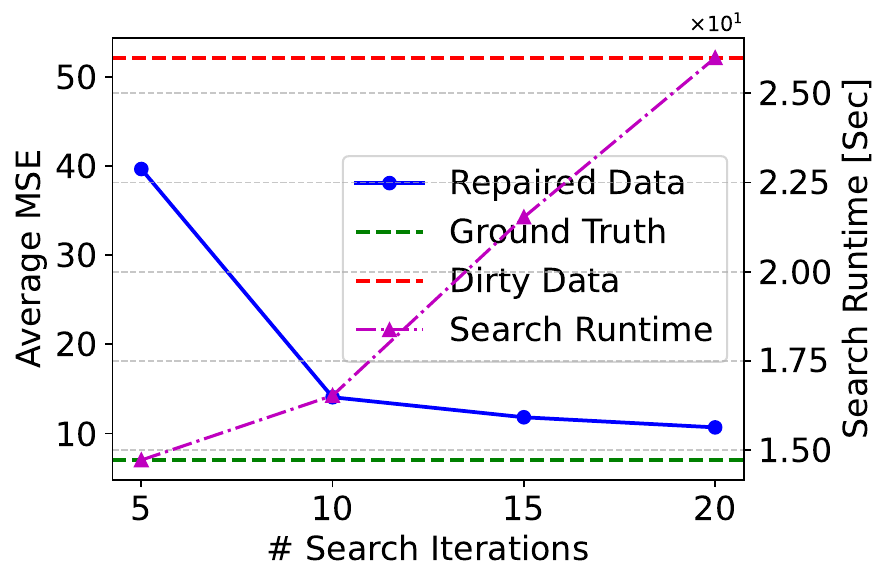}}	%
	\subfloat[Beers]{\label{fig:beers}\includegraphics[width=0.5\linewidth]{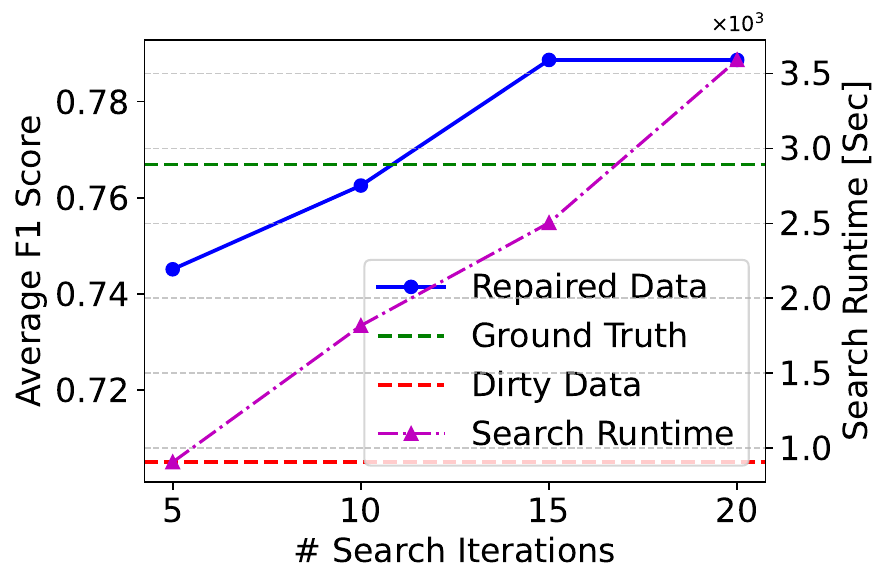}}
	\caption{Impact of the number of search iterations}
 	\label{fig:results} \vspace{-3mm}
 \end{figure}

Figure \ref{fig:results} presents a proof-of-concept evaluation of our iterative cleaning approach. We assess the impact of increasing search iterations on predictive performance, visualizing the results against baseline performances achieved using ground truth and dirty data. As an example, Figure \ref{fig:nasa} depicts the MSE of a decision tree model trained on data cleaned using the best tools identified by \PaperAcronym{} at each iteration count (ranging from 5 to 20). With fewer iterations, \PaperAcronym{} explores a limited subset of cleaning tools. However, as the iteration count increases, \PaperAcronym{} can evaluate a wider range of tool combinations, leading to the selection of more effective cleaning strategies. Notably, at 20 iterations, the model trained on data cleaned with the best-identified tools (Raha and ML Imputer in this instance) achieves an MSE of 10.7, closely approaching the performance achieved with the ground truth data. 

While achieving these results comes with a slight increase in search runtime (i.e., the time required for hyperparameter optimization), the iterative nature of our approach provides users with a powerful trade-off: they can directly control the balance between accuracy and runtime by adjusting the number of search iterations. Figure \ref{fig:beers} further illustrates the effectiveness of our approach on the Beers dataset for a multi-class classification task, demonstrating consistent performance improvements with increasing iterations. It is important to highlight that existing comparable methods like ActiveClean, BoostClean, and CPClean \cite{rein23} are restricted to binary classification tasks. In contrast, our iterative cleaning approach is broadly applicable to diverse ML tasks and model types, underscoring its versatility and potential impact.
%
\vspace{-3mm}\section{Reproducible Data Quality}\label{sec:reproducible_data_quality}
In this section, we present three features to facilitate the reproducibility of data quality experiments, including DataSheets, integration with mlflow, and integration with Delta Lake. \PaperAcronym{} allows users to generate DataSheets once error detection and repair tools have been executed on the dataset. These DataSheets compile an array of details about the dataset, including the dataset's name, locations for both the input dirty dataset and the repaired dataset, the shape (number of rows and columns) of the dataset, the detection tools applied, the number of erroneous cells identified in the dataset, the repair tools executed, and the configurations of such tools. It is important to mention that \PaperAcronym{} enables users to download the DataSheets and upload them later to reproduce the same data preparation steps. 

Aside from DataSheets, \PaperAcronym{} employs the MLflow library\footnote{https://mlflow.org/} to facilitate the tracking of parameters and artifacts. Each time an error detection or repair operation is executed, the specific parameters and artifacts are logged and locally stored. This enables users to retrieve and review the operational details at their convenience. To streamline access and management, the runs are segmented into distinct groups, referred to as \quotes{experiments}. These are specifically categorized under \quotes{Detection} and \quotes{Repair}, offering a systematic and intuitive way to navigate through the logged operations. This structured approach enhances the user experience and promotes efficient data operations. Compared to MLflow, DataSheets are persisted in the JSON format, providing a human-readable and highly interoperable data structure. 


Alongside experiment tracking, \PaperAcronym{} also monitors various versions of a dataset with the assistance of the Delta Lake library~\footnote{https://delta-io.github.io/delta-rs/python/}. This library is built upon the delta-rs Rust library, providing a robust foundation for dataset versioning. We opted for Delta Lake due to its simplicity in both setup and usage, particularly through its Python API. Unlike other data versioning libraries, Delta Lake does not necessitate preliminary setup requirements such as establishing an SQL connection, configuring a Kubernetes cluster, or initializing a Git repository. This straightforward approach reduces complexity and enhances user experience, making it an optimal choice for our dataset versioning needs.

Upon the initial upload of a dataset, a Delta Lake is instantiated. This Delta Lake essentially serves as a repository for the dataset, housing all the versions and transformations the dataset undergoes. The uploaded dataset is stored within this Delta Lake as a DeltaTable. A DeltaTable is a high-performance, format-agnostic, and schema-enforced collection of data, providing a structured and scalable framework for data storage. One of the key advantages of DeltaTable is its seamless interoperability with pandas DataFrames, facilitated by the methods provided by the Delta Lake library. It can be easily converted to a DataFrame for analysis and manipulation, and conversely, a DataFrame can be readily stored as a DeltaTable post-processing. This flexibility enhances the efficiency and versatility of data operations within \PaperAcronym{}.

If a DeltaTable already exists for a dataset from previous uploads, the user has the option to specify a version number during the dataset upload process. If the specified version number exists within the Delta Lake, the corresponding version of the dataset will be loaded for use. In scenarios where the indicated version does not exist, or the user does not provide a version number, the uploaded dataset will be stored as a new version within Delta Lake. Notably, this process does not overwrite or erase previous versions. Each iteration of the dataset is preserved within Delta Lake, maintaining a comprehensive record of dataset versions. This allows for historical tracking, comparison across versions, and the ability to revert to earlier versions if needed, thereby enhancing the robustness and flexibility of the data management system.

Once a user executes an error repair method, the resultant repaired dataset is stored within Delta Lake as a new, distinct version. This ensures that the dataset's progression through each error repair operation is precisely tracked, thereby maintaining a comprehensive record of the dataset's evolution. If a DataSheet is generated by the user, it will contain the version number of the dataset that has been used for error detection, as well as the version number of the dataset post-repair. This information provides a clear reference of the dataset's status at various stages of the error detection and repair process. By incorporating these version numbers into the datasheet, we enhance its utility as a comprehensive report of the data operations.

%
\section{Conclusion \& Future Work}\label{sec:conclusion}

This demonstration paper introduces \PaperAcronym{}, an interactive and ML-oriented dashboard aimed at streamlining and automating the management of tabular data quality. We elaborated on integrating a diverse set of data profiling, error detection, and repair tools through REST APIs, enhancing \PaperAcronym{}'s extensibility and functionality. Additionally, we described our iterative cleaning approach, which automatically selects the most effective cleaning tools based on the performance of downstream ML models, thus optimizing the data-cleaning pipeline. Moreover, we introduced how \PaperAcronym{} enables domain experts to contribute their knowledge for improved cleaning accuracy through supporting interactive rule validation, data labeling, and custom rule definition. While \PaperAcronym{} offers a comprehensive approach to data quality management, several avenues for future work exist: (1) Exploring more intuitive and user-friendly ways to interact with \PaperAcronym{}, such as natural language processing for rule definition and visual analytics for data exploration, can further enhance user experience. (2) Integrating explainability techniques into the error detection and repair process would give users insights into why specific errors were flagged and how corrections were made, fostering trust and understanding. (3) Enhancing the iterative cleaning module by incorporating more advanced hyperparameter optimization techniques and exploring the use of reinforcement learning for dynamic tool selection.
\bibliographystyle{ACM-Reference-Format}
\bibliography{main}
\end{document}